\documentclass[a4paper]{jpconf}
\usepackage{graphicx}

\begin{document}

\title{Investigating the Effects of Finite Resolution on Observed Transverse Jet Profiles}

\author{E. Murphy, D.C. Gabuzda}
\address{Physics Department, University College Cork, Cork, Ireland}
\ead{egm@student.ucc.ie}

\begin{abstract}
Both the emission properties and evolution of Active Galactic Nuclei (AGN) radio jets are dependent on the magnetic fields that thread them. Faraday Rotation gradients are a very important way of investigating these magnetic fields, and can provide information on the orientation and structure of the magnetic field in the immediate vicinity of the jet; for example, a toroidal or helical field component should give rise to a systematic gradient in the observed Faraday rotation across the jet, as well as characteristic intensity and polarization profiles. However, real observed radio images have finite resolution, usually expressed via convolution with a Gaussian beam whose size corresponds to the central lobe of the point source response function. This will tend to blur transverse structure in the jet profile, raising the question of how well resolved a jet must be in the transverse direction in order to reliably detect transverse structure associated with a helical jet magnetic field. We present  results of simulated intensity, polarization and Faraday rotation images designed to directly and empirically investigate the effect of finite resolution on observed transverse jet structures.
\end{abstract}

\section{Introduction}

At radio wavelengths the jets of active galaxies emit synchrotron radiation, which is characterised by appreciable linear polarization, with the plane of the polarization perpendicular to the plane of the jet magnetic field in the optically thin region. The polarization structure provides information about the structure of the magnetic fields threading these jets, which influence the evolution and emission properties of the jets, as well as their stability.  In addition,  an understanding of the magnetic field structure is important if we wish to correctly infer the intrinsic jet structure and physical processes occurring in the jet from the observed radio images. Yet despite much observational effort the nature of the magnetic field structures of AGN remain incompletely understood. Several types of observational results suggest that the magnetic field threading the jet may have a significant helical component on parsec scales. 

First, helical magnetic fields in a jet with perfect circular symmetry yield projections of the magnetic fields onto the plane of the sky which are either parallel or perpendicular to the jet axis, as is frequently observed [1,2]. Second, many parsec-scale jets show obvious antisymmetries in total intensity and polarization profiles that are reminiscent of those revealed in the helical field simulations of Laing [2,3]. Third, the presence of transverse Faraday rotation gradients across a number of VLBI jets can also be explained by helical fields in the immediate vicinity of the jets [4].

These observational results are all affected by the resolution of the observations used to detect them. In order to use these as identifiers for potential helical magnetic fields threading the jets of AGN, it is very important to understand the effects of finite resolution on observed transverse jet profiles.

\begin{figure}
\includegraphics[width=\columnwidth]{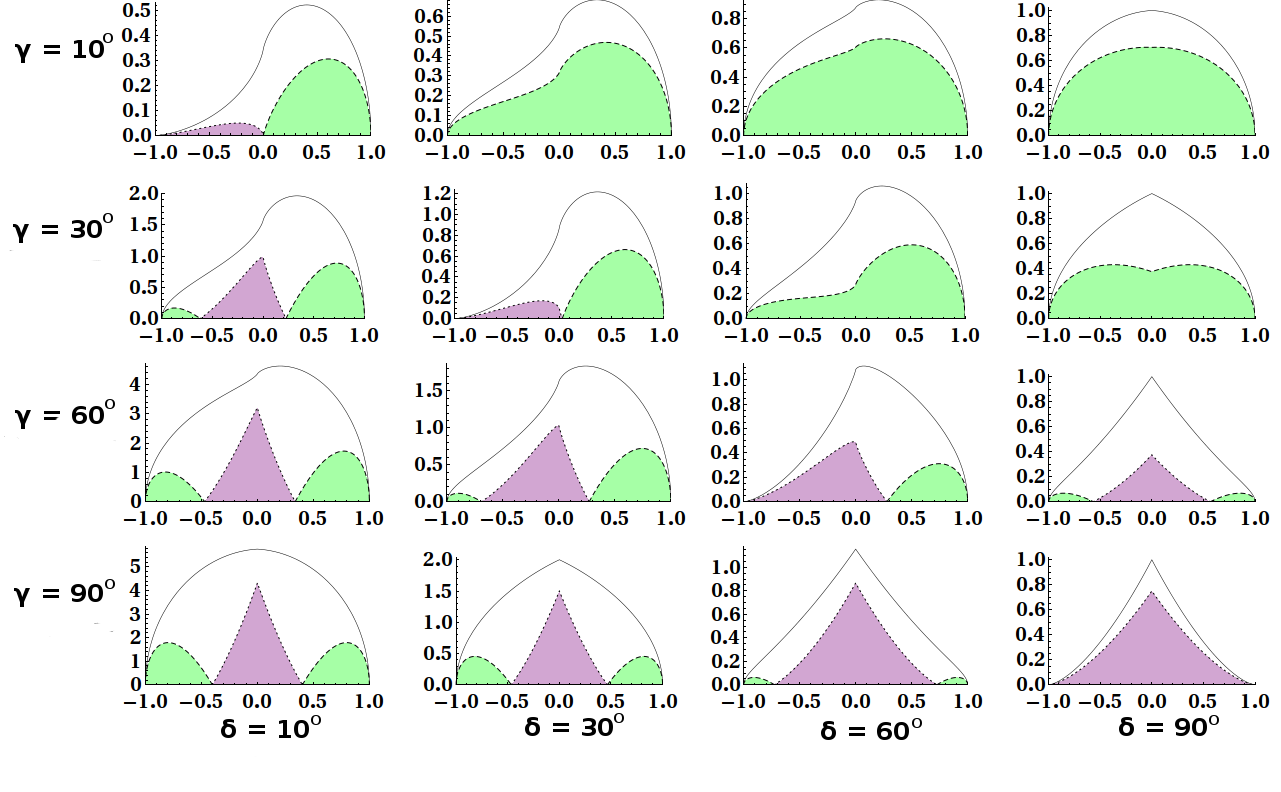}
\caption{Transverse structure produced by the helical field model, for various viewing angles, $\delta$, and helix pitch angles, $\gamma$. Solid lines correspond to total intensity, purple regions to longitudinal polarization and green regions to transverse polarization. The vertical scaling is arbitrary; the centre of the jet is at 0.0 on the horizontal axis.}
\label{Configs}
\end{figure}

\begin{figure}
\includegraphics[width=\columnwidth]{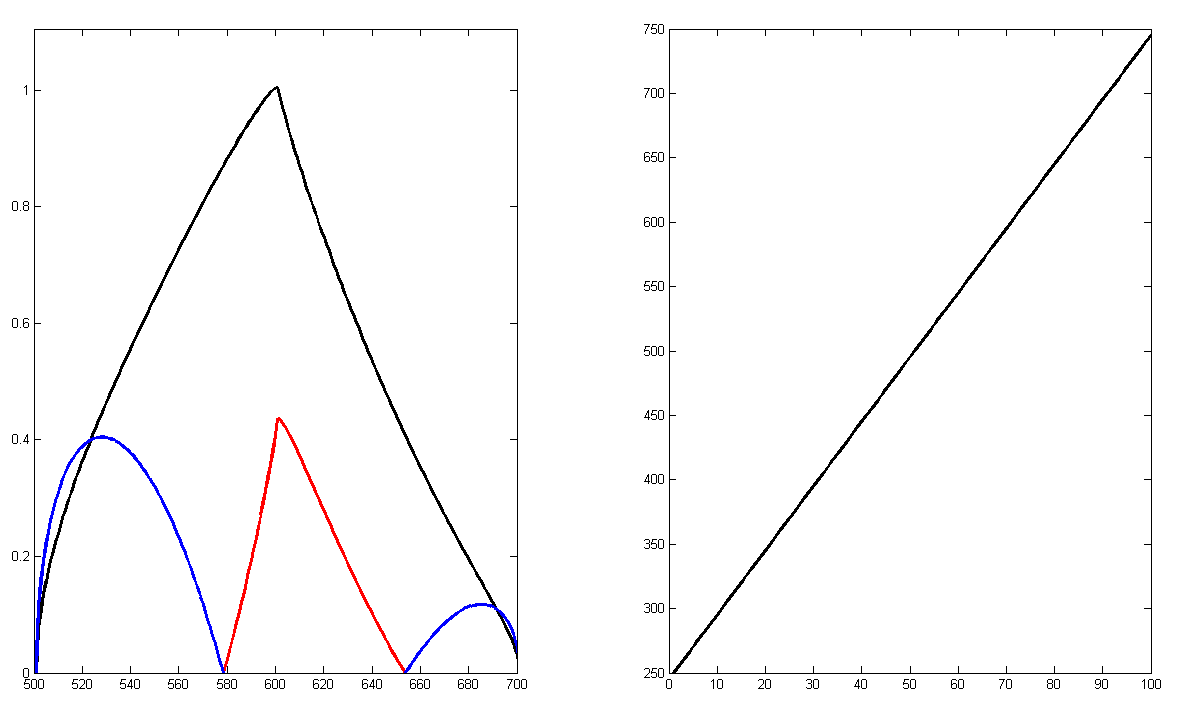}
\caption{Plots of the infinite resolution I and P (left) and Rotation Measure profiles (right) used  in analysis. Transverse profiles of I and P were generated for a jet with $\delta = 80^{\circ}$ and $\gamma = 53^{\circ}$. The black, blue and red lines correspond to total intensity, transverse polarization and longitudinal polarization, respectively. The I and P values are normalized to the peak I value. The polarization has in addition been scaled by a factor of two in order to increase visibility. The vertical axis in the RM plot is in rad/m$^2$. The horizontal scales are in arbitrary units: the intrinsic jet width is 200, the centre of the jet is at 600 in the left panel and at 50 in the right panel.}
\end{figure}

\section{Total Intensity and Polarized Intensity Profiles}

To investigate the effects of finite resolution on transverse jet profiles, infinite resolution intensity (I) and polarization (P) profiles where generated using a simple helical field model and then convolved with various beams. The model  assumes a helical B field of constant pitch angle and uniform flux density threading a cylindrical jet, and can be used to predict the I and P distributions across a jet as functions of the helical pitch angle ($\gamma$) and viewing angle ($\delta$) [2,3,5]. Both $\gamma$ and $\delta$ are in the rest frame of the jet. Fig.~\ref{Configs} shows the asymmetric intensity and polarization structure expected for various values of $\delta$ and $\gamma$. 

An example of the effects of convolution with a Gaussian beam on the I and P transverse profiles can be seen in Figs.~2 and 3. The infinite resolution I and P profiles are shown in Fig.~2 (left). Convolution with a Gaussian beam (Fig.~3) rapidly reduces the asymmetry in transverse I profiles, but reduces the asymmetry expected in P profiles much more slowly. One of the major effects convolution can have on transverse P profiles is changing the observed polarization configuration. In this example, the original polarization structure was `spine and sheath', a spine of longitudinal polarization with a sheath of transverse polarization (Fig.~2). When convolved with a beam whose width is 0.2 times the intrinsic jet width, the intensity profile appears only slightly asymmetric, but the asymmetric `spine and sheath' polarization structure is clearly visible (Fig.~3, top left). After convolution with a beam of width equal to 0.6 times the intrinsic jet width, the intensity profile has become very symmetric and one sheath is no longer observed (Fig.~3, top right). The P profile for this case is now longitudinal on one side, transverse on the other. After convolution with a beam of width equal to the intrinsic jet width, the longitudinal `spine' becomes very weak (Fig.~3, bottom left). Finally, after convolution with a beam width equal to 1.4 times the intrinsic jet width, only longitudinal polarization offset to one side of the jet is observable (Fig.~3, bottom right). However, even after convolution with a very large beam such as this last one, the position of the maximum polarization intensity is still clearly offset from the position of maximum total intensity. Thus, the observed transverse P structure can still reflect the presence of helical magnetic fields threading the jets of AGN even if the beam used in the observations is large with respect to the size of the jet, although it will not be possible in this case to determine accurate values of $\gamma$ and $\delta$ based on the observed images.

\begin{figure}
\includegraphics[width=\columnwidth]{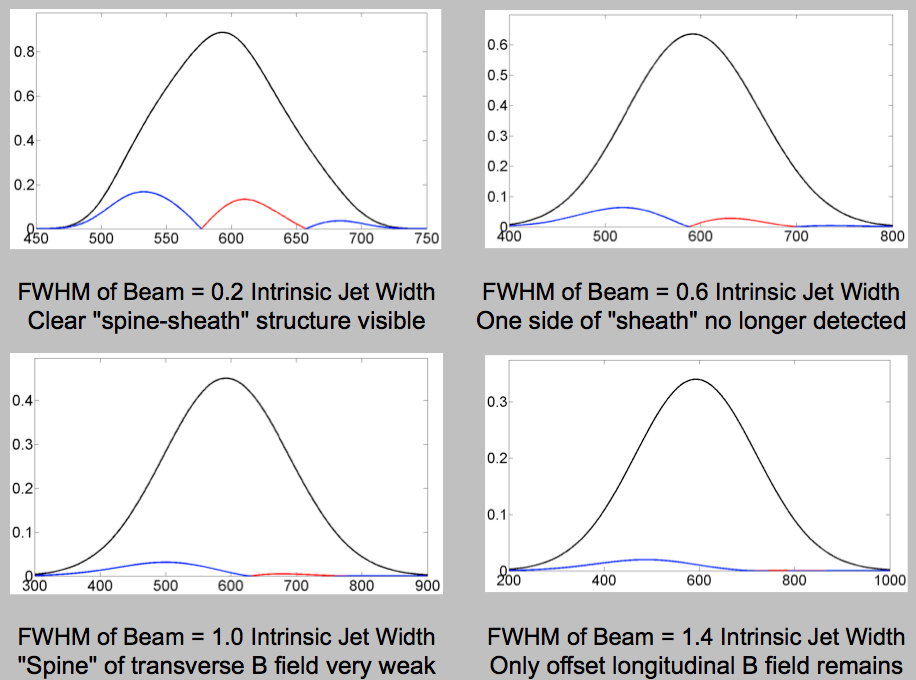}
\caption{Transverse profiles of I and P generated for a jet with $\delta = 80^{\circ}$ and $\gamma = 53^{\circ}$ after convolution with Gaussian beams of various sizes. The black, blue and red lines correspond to total intensity, transverse polarization and longitudinal polarization, respectively. The I and P values are normalized to the peak I value. The polarization has in addition been scaled by a factor of two in order to increase visibility. The horizontal scale is in arbitrary units, with an intrinsic jet width of 200 with a centre at 600. }
\label{TransI}
\end{figure}

\begin{figure}
\includegraphics[width=\columnwidth]{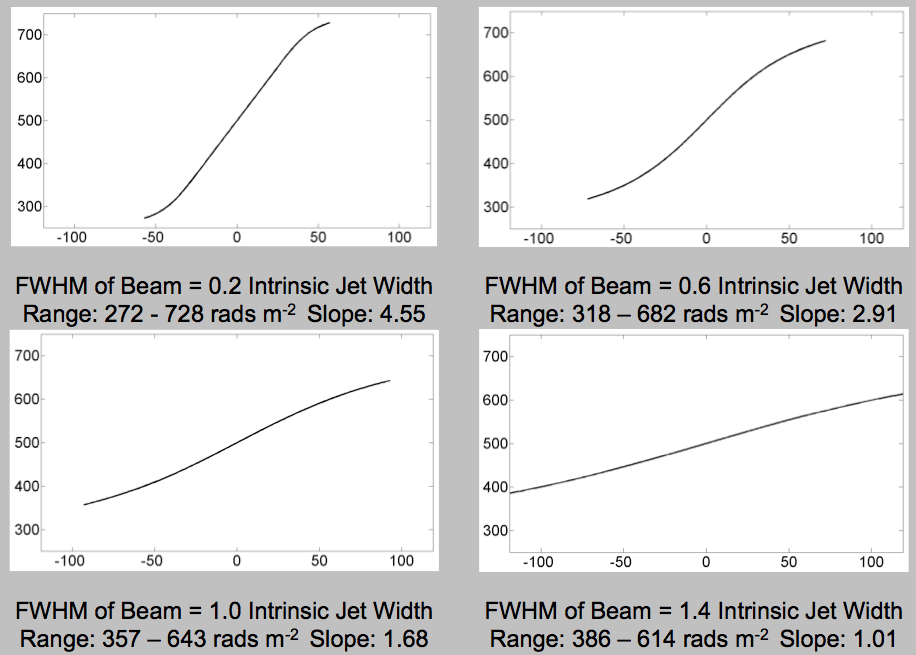}
\caption{Transverse profiles of Faraday Rotation Measure after convolution with Gaussian beams of various sizes. The vertical axes are in rad/m$^2$. The values given for the slopes correspond to the slope of the best fit line fitted to the gradients. The horizontal scale is arbitrary, with a intrinsic jet width of 100 and the jet centre at 0.}
\label{TransRM}
\end{figure}

\section{Faraday Rotation}

Faraday Rotation occurs when an electromagnetic wave propagates through a  region with charged plasma and a magnetic field. Faraday Rotation rotates the polarization of the electromagnetic wave because the left-hand circularly polarized component of the EM wave has different refractive index than the right-hand circularly polarized component, given by
\begin{equation}
\chi=\chi_0 + RM \lambda^2 
\end{equation}
where $\chi$ is the observed polarization angle, $\chi_0$ is the emitted polarization angle, $\lambda$ is the wavelength and RM is the Rotation Measure. The Rotation Measure is given by
\begin{equation}
RM=\frac{e^3}{8 \pi^2 \epsilon_{0} m_{e}^2 c^3} \int n_e \overrightarrow{B} \bullet \overrightarrow{dl}
\end{equation}
where e is the elementary charge, $\epsilon_0$ is the permittivity of free space, $m_{e}$ is the mass of the electron, $n_e$ is the number density of electrons in the plasma and B is the magnetic field strength in the plasma. We would expect to observe a transverse RM gradient if a helical B field threads the jet, due to the systematic change in $\overrightarrow{B}$ $\bullet$ $\overrightarrow{dl}$ across the jet[4].\newline\newline

In order to analyse the effects of finite resolution on observations of a RM gradient, the effects of a calculated RM gradient produced by the helical field model were applied to the Stokes Q and U profiles for an infinite resolution jet. The resulting Q and U profiles were then convolved with various beams, and the RM profile was determined from these convolved Q and U profiles. The original RM gradient applied to the Q and U profiles ranged from 250 to 750 rad/m$^{2}$ and had a slope of 5 (Fig.~2). Figure \ref{TransRM} shows the observed RM gradients after convolution with the same beams used in Figure \ref{TransI}.  Convolution significantly changes the values of the observed Faraday RMs  but does not destroy the RM gradients produced by the helical field, essentially independent of the size of the beam. Transverse RM gradients can potentially remain visible, even with beams significantly larger than the width of the jet.

\section{Conclusion}

The presence of a helical magnetic field threading an AGN jet can be inferred from characteristic transverse structures in the intensity, polarized intensity and rotation measure profiles. The detection of this transverse structure is affected by the resolution of the observations used. Convolution with  a Gaussian beam rapidly reduces the asymmetry in the I profiles, while weakening the P structure much less severely. Appreciable transverse P structure remains visible even when the I profile has become completely symmetrical, and even when the beam is appreciably larger than the intrinsic jet width. Thus, it may be possible to clearly detect transverse structure consistent with a helical magnetic field, but not  reliably deduce the parameters of the field if the beam is large. Convolution also significantly changes the values of the Faraday RMs observed across AGN jets, but does not destroy the RM gradients produced by the helical field, even in the case of large beams. Here as well, the presence of the RM gradient associated with a helical B field is robust, but it is not possible to accurately deduce the parameters of the field if the beam is large compared to the jet. This suggests that the proposed criterion of [6] that a transverse RM gradient must span three beamwidths to be reliable is inappropriate and overly conservative. The important criteria are the others proposed in [6]: that the candidate RM gradients be monotonic and smooth, that the RM change by at least three times the error across the gradient, and that the region sampled be predominantly optically thin at all the frequencies used to make the RM distribution. These three criteria have, in fact, been satisfied by the majority of published RM analyses reporting the presence of transverse RM gradients [7--12].

\section{Acknowledgments}

Funding for this research was provided by the Irish Research Council for Science Engineering and Technology (IRCSET).

\end{document}